\documentclass[twocolumn,superscriptaddress,showpacs]{revtex4-1} 

\usepackage{graphicx} 
\usepackage{natbib} 
\usepackage[usenames,dvipsnames]{color} 
\usepackage{amsmath} 
\usepackage{amssymb} 
\usepackage{soul} 
\usepackage{ifthen} 
\usepackage{enumitem}
\usepackage{upgreek}

\newcommand{\rt}[1]{\sqrt{#1}}
\newcommand{\bra}[1]{\langle{#1}|}
\newcommand{\ket}[1]{|{#1}\rangle}

\def\be#1\ee{\begin{equation}#1\end{equation}}
\def\ba#1\ea{\begin{align}#1\end{align}}
\def\bg#1\eg{\begin{gather}#1\end{gather}}
\def\t{\text}

\newcommand{\abs}[1]{\lvert#1\rvert}

\def\shownote{0} 
\newcommand{\note}[1]{\ifthenelse{\shownote=1}{\textcolor{Red}{[[#1]]}}{}}
\def\showaddmat{0} 
\newcommand{\addmat}[1]{\ifthenelse{\showaddmat=1}{\textcolor{Gray}{[[#1]]}}{}}

\def\shownal{0} 
\newcommand{\nal}[1]{\ifthenelse{\shownal=1}{\textcolor{YellowOrange}{[[#1]]}}{}}

\def\shownmc{0} 
\newcommand{\nmc}[1]{\ifthenelse{\shownmc=1}{\textcolor{Brown}{[[#1]]}}{}}

\begin{document}

\title{Quadrature readout and generation of squeezed states of a harmonic oscillator using a qubit-based indirect measurement}

\author{M. Canturk}
\affiliation{Institute for Quantum Computing, Department of Mechanical and Mechatronics Engineering, and Waterloo Institute for Nanotechnology, University of Waterloo, Waterloo, ON, Canada N2L 3G1}

\author{A. Lupascu \footnotemark[1] \footnotetext[1]{Corresponding author: adrian.lupascu@uwaterloo.ca}}
\affiliation{Institute for Quantum Computing, Department of Physics
and Astronomy, and Waterloo Institute for Nanotechnology, University
of Waterloo, Waterloo, ON, Canada N2L 3G1}
\email{adrian.lupascu@uwaterloo.ca}

\date{ \today}

\begin{abstract}
We present a protocol for measuring the quadrature of a harmonic oscillator (HO). The HO is coupled to a qubit, with an interaction modulated by the qubit control and effectively proportional to the HO quadrature $I$. Repeated measurement of the qubit leads to gradually increasing information on the quadrature $I$, leading to squeezing. We derive an analytical formula for the quadrature variance, $(\Delta I)^2 = 1/(1+4\phi^2 s)$, with $\phi$ the product of interaction strength and interaction time and $s$ the number of repetitions of the measurement. We discuss the robustness of this scheme against decoherence. We find that this protocol could lead to significant squeezing in a realistic setup formed of a superconducting flux qubit used to measure an electrical or mechanical resonator.

\end{abstract}
\maketitle

\textit{Introduction.---} The quadratures of a quantum harmonic oscillator (HO) are operators defined as $I= a + a^\dag $ and $Q=-i(a - a^\dag)$, with $a$ ($a^\dag$) the HO annihilation (creation) operator. The variances for these operators are constrained by the uncertainty principle, which imposes $(\Delta I)^2 \, (\Delta Q)^2 \geq 1$. Squeezed states are characterized by a variance in one quadrature reduced below 1 at the expense of increased uncertainty in the other quadrature. Quadratures are constants of motion for a HO, which allows, in principle, their high precision measurement using a quantum non-demolition readout~\cite{braginsky_quantum_1992}. Therefore, by monitoring one of the two quadratures, a signal acting on the HO can be detected with a precision only limited by the ability to prepare the chosen quadrature in a low uncertainty state, making quadratures useful for sensitive detection~\cite{caves_measurement_1980}. Squeezed states have applications also in quantum measurements~\cite{didier_fast_2015} and quantum information based on continuous variables~\cite{lloyd_quantum_1999}.

Recently, developments in the field of control of mechanical resonators have led to the experimental demonstration of preparation and detection of squeezed states~\cite{wollman_quantum_2015,pirkkalainen_squeezing_2015,lei_quantum_2016,lecocq_quantum_2015,lei_quantum_2016}. In the field of superconducting circuits, squeezed states of superconducting electromagnetic resonators have become an essential ingredient in quantum limited amplifiers (see \emph{e.g.} Ref.~\cite{castellanos-beltran_amplification_2008}). Various methods have been proposed to implement squeezing in mechanical systems, including back-action evading schemes based on two-tone driving~\cite{clerk_back-action_2008}, engineered dissipation~\cite{kronwald_arbitrarily_2013}, parametric driving~\cite{szorkovszky_mechanical_2011,vinante_feedback-enhanced_2013}, stroboscopic measurements~\cite{ruskov_squeezing_2005}, pulsed optomechanics~\cite{vanner_pulsed_2011}, and squeezed light injection~\cite{jahne_cavity-assisted_2009}. In superconducting electromagnetic resonators, squeezing relies on non-linearities due to Josephson junctions and parametric amplification~\cite{everitt_superconducting_2004,zagoskin_controlled_2008}. Nevertheless, finding versatile and efficient methods to generate squeezed states remains a topic of growing importance.

In this paper, we present a method to perform high fidelity quadrature measurements and generate squeezed states of a HO. The HO interacts with the qubit via a $(a+a^\dag)\sigma_z$ interaction, where $\sigma_z$ is a Pauli operator in the qubit energy eigenbasis. The qubit is controlled with resonant pulses, used to induce transitions between its energy eigenstates, separated by half the period of the HO. A superposition of qubit energy eigenstates acquires a phase, dependent on the quadrature $I$, which is detected in a Ramsey-type experiment. We show how repetition of this sequence leads to increasing information on the quadrature $I$ and a corresponding reduction in the uncertainty $\Delta I $ corresponding to squeezing.  We discuss the application of this protocol to measurement of superconducting electromagnetic resonators and nano-mechanical resonators, taking into account non-idealities including decoherence and qubit detection errors. We note that our proposed scheme involves an effective modulation of the interaction between the HO and the qubit detector, bearing a connection with the generic modulation scheme of Thorne \emph{et al.}~\cite{thorne_quantum_1978}. The periodic interaction has similarities with stroboscopic measurements~\cite{ruskov_squeezing_2005}, with one important difference being that the interaction is continuous, leading to increased coupling strength. The same qubit control pulse scheme was proposed for ac-magnetic field coherent~\cite{taylor_high-sensitivity_2008} and incoherent~\cite{kolkowitz_coherent_2012,bennett_measuring_2012} detection and shown to be amenable to classical quadrature measurements~\cite{Bal_2012_QubitDetector}. In Ref.~\cite{rao_heralded_2016}, a similarly modulated interaction is used for heralded cooling and squeezing. In marked contrast with Ref.~\cite{rao_heralded_2016}, the choice we take for qubit detection implements quadrature measurement and leads to generation of low variance states for any measurement result.

\begin{figure}[!]
\includegraphics[width=86mm]{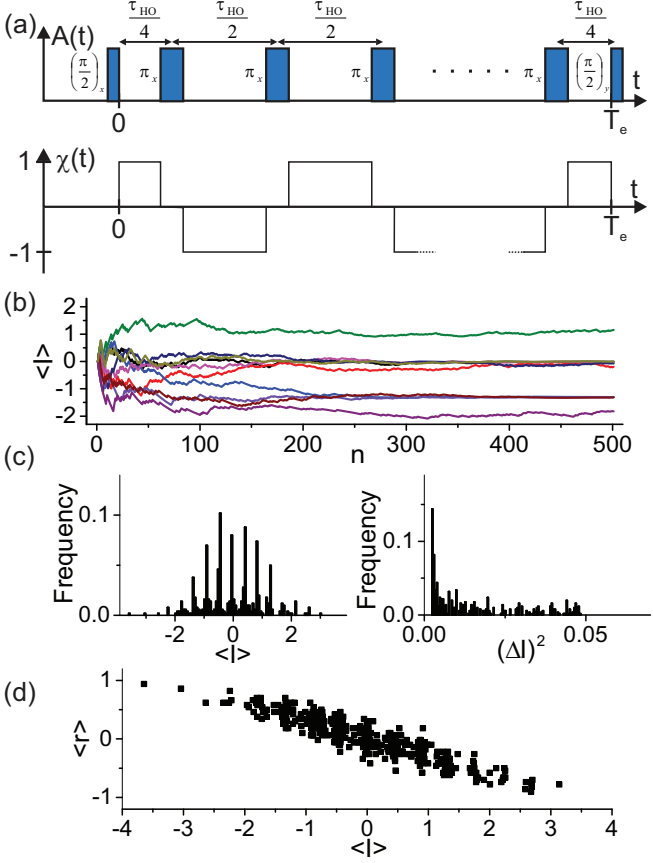}
\caption{\label{fig1}(color online). (a) Pulse sequence (top) used to control the qubit and modulation function for the qubit-HO interaction (bottom). (b) The quadrature average $\langle I \rangle$ as a function of the measurement step $n$, for a set of simulated trajectories. (c) The distribution of $\langle I \rangle$ and $(\Delta I)^2$ after $s=500$ measurement steps, extracted from 500 trajectories. (d) The average of qubit readout results over the last 50 points in each measurement trajectory with $s=500$, versus $\langle I \rangle$. In all the simulations $\phi=0.159$.}
\end{figure}

\textit{Measurement protocol.---} We consider a system formed of a HO coupled to a qubit, with the Hamiltonian $H = H_{\t{HO}} + H_{\t{qb}} + H_{\t{qb,c}} + H_{\t{int}}$. We have $H_{\t{HO}} = \omega_\t{r} a^\dag a$, $H_{\t{qb}} = -\frac{\omega_\t{ge}}{2}\sigma_z$, $ H_{\t{qb,c}} = f(t)\sigma_x$, and $H_{\t{int}} = g (a + a^\dag)\sigma_z$, with $\omega_\t{r}$ the HO resonance frequency, $\sigma_z$ ($\sigma_x$) Pauli z(x) operators in the qubit energy eigenbasis, $\omega_\t{ge}$ the qubit transition frequency, $f(t)$ a qubit control term, and $g$ the HO-qubit coupling strength. The qubit is controlled with resonant pulses, \emph{i.e.} by setting $f(t) = A(t) \cos (\omega_\t{ge} t + \varphi(t))$, with the amplitude $A(t)$ and the phase $\varphi(t)$ changing slowly as a function of time. We make a transformation to a rotating frame, described by the unitary operator $U_\t{rf} = e^{i(H_\t{HO} + H_{\t{qb}})t}$. In this frame the Hamiltonian is $H_\t{rf} = g(a e^{-i\omega_r t} + a^\dag e^{i\omega_r t})\sigma_z + \frac{A(t)\cos \varphi (t)}{2} \sigma_x - \frac{A(t)\sin \varphi (t)}{2} \sigma_y$, where we used the rotating wave approximation.

\begin{figure}[!]
\includegraphics[width=86mm]{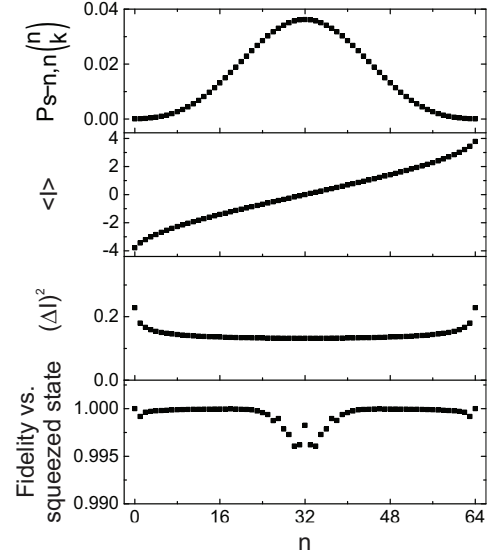}
\caption{\label{fig2} Top three panels: the probability of a measurement sequence with $n$ results $r=1$, the quadrature average, and the quadrature variance, respectively, versus measurement step $n$ obtained from Eqs.~\ref{eq:probsofn}, \ref{eq:Iavn}, and \ref{eq:I2avn}. Bottom panel: fidelity against a squeezed state versus $n$. We take $\phi=0.159$ and $s=64$.}
\end{figure}

The measurement protocol consists of repeating the procedure shown schematically in Fig.~\ref{fig1}(a). The qubit and the HO start in a separable state $\ket{g} \otimes \ket{\alpha}$, where $\ket{g}$ ($\ket{e}$) is the qubit ground (excited) state and $\ket{\alpha}$ is an arbitrary HO state. Next, the qubit is controlled using a Carr-Purcell-Meiboom-Gill type control sequence~\cite{meiboom_modified_1958}, consisting of the pulses $\left(\frac{\pi}{2}\right)_x - \left[ \left(\pi\right)_x \right]^{N_\t{p}}  -\left(\frac{\pi}{2}\right)_y$, as shown schematically in Fig.~\ref{fig1}(a). Each rotation $\theta_\beta$ is a rotation of angle $\theta$ around axis $\beta = x$ or $y$. The first control pulse changes the qubit state to $\frac{1}{\rt{2}}(\ket{g} - i \ket{e})$. The evolution of the combined system during the time interval between the initial and final pulses is given by the unitary operator $U_e = U_c \mathcal{T}\exp{\left(-i \int_{0}^{T_\t{e}}dt\, H_\t{eff}(t)\right)}$, where the effective Hamiltonian $H_\t{eff} (t) = g \chi (t)  (a e^{-i\omega_r t} + a^\dag e^{i\omega_r t})\sigma_z $ and $U_c = I_\t{qb}$, the identity operator for the qubit, for for $N_p$ even, and $U_c = e^{-i\pi/2 \sigma_x}$ for $N_p$ odd. After the final pulse, the qubit is measured projectively, and the measurement result $r=1$ ($-1$), corresponding to projection in the excited (ground) state, is recorded. Following measurement, the qubit is reset to its ground state, in preparation for the next repetition.

The evolution of the coupled qubit-HO system between the two $\pi/2$ pulses in Fig.~\ref{fig1}(a) is exactly described by the Hamiltonian $H_\t{avg} = \frac{2}{\pi} g \sigma_z I$, with the quadrature $I = (a+a^\dag)$, obtained by averaging the Hamiltonian $H_\t{eff} (t)$ over the complete duration of the interaction. Qualitatively speaking, the qubit superposition $\frac{1}{\rt{2}}(\ket{g} - i \ket{e})$ prepared by the first $\pi/2$ pulse acquires a phase that depends on the quadrature $I$. The combination of the $(\pi/2)_y$ pulse and measurement in the energy eigenbasis constitutes a measurement in the $\sigma_x$ eigenbasis, which provides information on the quadrature $I$.

\textit{Analysis of the measurement process.---} Next, we present an analysis of the measurement process. We consider a series of $s$ repetitions of the protocol illustrated in Fig.~\ref{fig1}(a). For repetition $i$ ($i=\overline{1,s}$), the starting state of the combined system is $\ket{g}\otimes \ket{\alpha_{i-1}}$. After interaction and immediately prior to measurement, the state becomes $\ket{g} \otimes D_\t{g}\ket{\alpha_{i-1}} + \ket{e} \otimes D_\t{e}\ket{\alpha_{i-1}}$, where $D_\t{g}=-\frac{1}{2}(D-iD^\dag)$ and $D_\t{e}=-\frac{1}{2}(D+iD^\dag)$. Here $D=\mathcal{D}(i\phi)$, with $\mathcal{D}(\beta) = e^{\beta a^\dag - \beta^* a}$ the displacement operator of amplitude $\beta$~\cite{walls_1995_1}, and $\phi = \frac{2}{\pi}g T_e$. The measurement result $r=-1$ ($1$) occurs with probability $P_g = || D_\t{g}\ket{\alpha_{i-1}} ||^2$ ($P_e = || D_\t{e}\ket{\alpha_{i-1}} ||^2$) and induces a post-measurement state $\ket{\alpha_i} = D_\t{g}\ket{\alpha_{i-1}} / \rt{P_g}$ ($D_\t{e}\ket{\alpha_{i-1}} / \rt{P_e}$). By iteration, the probability to obtain a set of measurements such that $n$ of the $s$ results are $+1$, is given by $P_{(s-n,n)} = || D_e^n D_g^{s-n} \ket{\alpha_0} ||^2$ and the resulting state is $D_e^n D_g^{s-n} \ket{\alpha_0}/\rt{P_{(s-n,n)}}$. We note that the probability and the conditioned state are independent of the order in which the $n$ results of value $1$ are obtained, due to $\left[ D_g,D_e \right] = 0$.

We first analyze the measurement action by stochastic numerical simulations. The HO is prepared in its vacuum state. We simulate a set of measurement sequences, each consisting of $s$ measurements. Within each sequence, we assign at each step a measurement result $r$, by drawing the random number $r=1$ ($-1$) with probability $P_e$ ($P_g$), and we also assign the corresponding conditioned state. In Fig.~\ref{fig1}(b) we show, within each sequence, the evolution of the average quadrature $\langle I \rangle$ versus the measurement step. We observe that after undergoing fluctuations, $\langle I \rangle$ settles to a nearly constant value. In Fig.~\ref{fig1}(c) we show the histogram of the average $\langle I \rangle$ and of the variance $(\Delta I)^2$. The distribution of $\langle I \rangle$ is consistent with the initial state probability, whereas $\Delta I$ is significantly reduced compared to the initial distribution. These features are a consequence of the quantum non-demolition type of interaction. Remarkably, the values taken by $\langle I\rangle$ are discrete, a feature that reflects the discrete nature of the information acquired from binary qubit readout results. In Fig.~\ref{fig1}(d) we show the average of the last few measurement results versus the final $\langle I \rangle$ for each sequence. The strong correlation demonstrates that the qubit readout is a suitable meter for the quadrature $I$. The results in Fig.~\ref{fig1} correspond to $\phi=0.159$. We observe similar results for preparation for other values of $\phi$, with a general tendency for $\langle I \rangle$ to converge faster and for $(\Delta I)^2$ to decrease as $\phi$ increases. We also observe similar results when the HO is prepared in coherent or thermal states.

We discuss next the properties of the measurement conditioned states. We consider the case in which the initial state of the HO is a coherent state of amplitude $\alpha_0$. The probability to detect the result $r=1$ for $n$ times out of $s$ repetitions, the corresponding average, and the corresponding average of the square of the quadrature are given respectively by
\begin{widetext}
\begin{equation}\label{eq:probsofn}
P_{(s-n,n)} = \frac{\big(-1\big)^{\frac{s}{2}-n}e^{-2\Re\{\alpha_0\}^2}}{2^{2s}} \sum_{k=0}^{2(s-n)}\sum_{\ell=0}^{2n}
{2(s-n)\choose k}{2n\choose\ell} i^{(\ell-k)} e^{+2\big(\Re\{\alpha_0\}+i\phi(s-k-\ell)\big)^2},
\end{equation}

\begin{equation}\label{eq:Iavn}
\left\langle I\right\rangle_{(s-n,n)}  = \frac{\big(-1\big)^{\frac{s}{2}-n}e^{-2\Re\{\alpha_0\}^2}}{2^{2s}P_{(s-n,n)}}
\sum_{k=0}^{2(s-n)} \sum_{\ell=0}^{2n}  {2(s-n)\choose k}{2n\choose \ell} i^{\ell-k} 2\bigg(\Re\{\alpha_0\}+i\phi(s-k-\ell)\bigg)
e^{+2\big(\Re\{\alpha_0\}+i\phi(s-k-\ell)\big)^2},
\end{equation}
and
\begin{equation}\label{eq:I2avn}
\left\langle I^2\right\rangle_{(s-n,n)} = \frac{\big(-1\big)^{\frac{s}{2}-n}e^{-2\Re\{\alpha_0\}^2}}
{2^{2s}P_{(s-n,n)}}
\sum_{k=0}^{2(s-n)} \sum_{\ell=0}^{2n}  {2(s-n)\choose k}{2n\choose \ell} i^{\ell-k} \bigg(1+4\big(\Re\{\alpha_0\}+i\phi(s-k-\ell)\big)^2\bigg)
e^{+2\big(\Re\{\alpha_0\}+i\phi(s-k-\ell)\big)^2}
\end{equation}
\end{widetext}
(see~\cite{si}).
Using these expressions, we calculate and show in Fig.~\ref{fig2} the probability for each result, which is given by $P_{(s-n,n)}$ multiplied by the combinatorial factor ${s\choose n}$, the average, and the variance versus $n$. These results show that measurement conditioned states have reduced variance in the quadrature $I$. It is interesting to consider whether the resulting states are squeezed states, as generated by a squeezing operator $\mathcal{S}(\epsilon) = \exp{ \left( \frac{\epsilon ^*}{2}a^2 - \frac{\epsilon}{2}{a^\dag}^2 \right) }$~\cite{walls_1995_1}. In Fig.~\ref{fig2} we also show the fidelity of the measurement conditioned state with respect to the state $\mathcal{D}(\langle I\rangle)_{s-n,n}\mathcal{S}(-\log (\Delta I)_{s-n,n})\ket{0}$. We find that, besides having reduced variance, the states prepared by measurement have a very high fidelity with respect to states generated by the squeezing operator.

\begin{figure}[!]
\includegraphics[width=86mm]{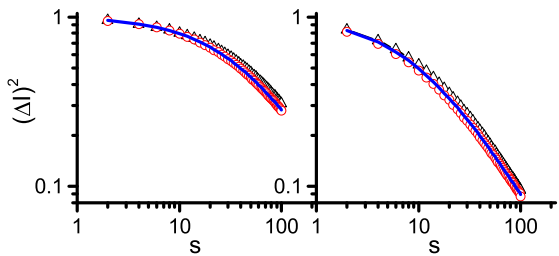}
\caption{\label{fig3}(color online). Average variance (triangles) and variance for the symmetric measurement ($n=s/2$) (dots) versus the number of measurement steps $s$. The solid line is the approximation in Eq.~\ref{eq:approxvariance}. The left and right panels correspond to $\phi=0.08$ and $\phi=0.159$ respectively.}
\end{figure}

We next consider the dependence of the variance on the number of measurement steps. For an initial vacuum state, the variance of the most likely state ($n=s/2$) as well as its average weighted over the probabilities of resulting states is shown in Fig.~\ref{fig3} for two values of $\phi$. Based on equations \ref{eq:probsofn},\ref{eq:Iavn}, and \ref{eq:I2avn}, we derived an analytical approximation for the variance~\cite{si},
\begin{equation}\label{eq:approxvariance}
(\Delta I)^2_{s-n,n} = 1/(1 + 4 \phi^2 s),
\end{equation}
which is in excellent agreement with the exact calculations, as shown in Fig.~\ref{fig3}.

\begin{figure}[!]
\includegraphics[width=75mm]{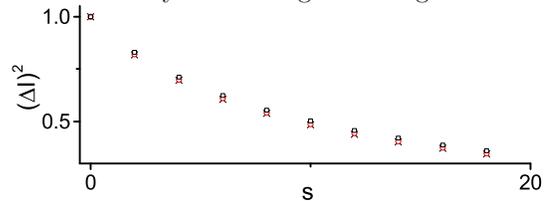}
\caption{\label{fig4}(color online). Variance without noise (crosses) and with noise, with a spectral density $A_\omega=(1.2\times 10^7\,\t{rad/s})^2$ (empty dots) and $A_\omega=(2.4\times 10^7\,\t{rad/s})^2$ (empty squares) versus the number of measurements.}
\end{figure}

\textit{The role of qubit dephasing.---} Given the fact that quadrature measurement relies on the detection of the phase of a qubit superposition, dephasing of a qubit induced by its environment should be considered. In the presence of dephasing, the projection operators $D_{\t{g}(\t{e})}$ become $D_{\t{g}(\t{e})}=-\frac{1}{2}(D-(+)e^{i\tilde{\phi}}iD^\dag)$, where $\tilde{\phi}$ is a random phase acquired by the qubit due to noise. The state conditioned by a given series of measurement results $r_1$, $r_2$,...,$r_s$ becomes a density matrix when averaged over noise realizations, and is given by
\begin{equation}
\rho_\mathbf{r}=\frac{1}{2^{2s}}\sum_{\mathbf{q_1},\mathbf{q_2}}i^{\mathbf{r}(\mathbf{q_1}-\mathbf{q_2})}C_{\mathbf{q_1},\mathbf{q_2}}D^{s-t_1}\ket{\alpha}\bra{\alpha}{D^\dag}^{s-t_2},
\end{equation}
where $\mathbf{q_1}$ and $\mathbf{q_2}$ are vectors of length $s$ with components $0$ or $1$, $t_{1(2)}$ is the sum of the components of $\mathbf{q_{1(2)}}$, and $C_{\mathbf{q_1},\mathbf{q_2}}=\exp{\left( -\frac{1}{2} (\mathbf{q_1}-\mathbf{q_2}) \mathbf{\widetilde{W}}(\mathbf{q_1}-\mathbf{q_2})   \right)}$, with $\mathbf{\widetilde{W}}$ the correlation matrix for the noise $\tilde{\phi}$. We considered quadrature measurement with $g=1$~MHz, $\omega_\t{r}=2\pi\times200$~MHz, $N_\t{p}=50$, and noise in the qubit frequency $\omega_{ge}$ with a spectral density $A_\omega/\abs{\omega}$. This type of noise spectral density is typical in superconducting qubits~\cite{bylander_2011_noisePCQ}; we take a typical value $A_\omega=(1.2\times 10^7\,\t{rad/s})^2$. A comparison of the variance without and with dephasing is shown in Fig.~\ref{fig4}. This level of noise produces a negligible effect on quadrature squeezing up to $s=$18. Even with significantly larger noise, $A_\omega=(2.4\times 10^7\,\t{rad/s})^2$, squeezing is degraded by less than 5~\%.

\textit{Experimental implementation.---} We briefly discuss the prospects for experimental implementation. We consider a superconducting flux qubit used to measure either an electrical or a mechanical resonator. The flux qubit has $\omega_{ge}=2\pi\times 10.8$~GHz, an energy level splitting at the symmetry point $\Delta= 2 \pi \times 4\,\t{GHz}$, a persistent current  $I_\t{p}=300$~nA, an energy relaxation time $T_{\t{1,qb}}=10\,\mu$s, an effective temperature $T_\t{qb}=50$~mK, and is subjected to intrinsic flux noise with a spectral density $A_\Phi/\abs{\omega}$ with $A_\Phi = 1\,(\upmu \Phi_0)^2$~\cite{orgiazzi_2016_FluxQubitsPlanar,Stern_2014_3DcavityFluxQubits,Yan_2016_FluxQubitRevisited}. The HO has $\omega_\t{r}=200$~MHz, a quality factor $Q=10,000$, and a temperature $T_\t{HO}=15$~mK. A coupling strength $g=2\pi\times 2$~MHz is achievable by inductive coupling of an electrical superconducting resonator or by embedding a moving beam into the qubit arm, similarly to the superconducting interferometer setup in Ref~\cite{etaki_motion_2008}. With the numbers given above, we find that a HO initially in its thermal state can be brought into a squeezed state with a variance reaching $(\Delta I)^2 = 0.4$. We note that the assumed value of $g$ is conservative. Larger coupling of the qubit to a mechanical HO is envisioned with optimized setups and coupling to an electrical HO can be straightforwardly be made over an order of magnitude larger than considered, leading to larger squeezing. We expect that further optimization of other parameters of the measurement protocol will also result in increased squeezing.

\textit{Conclusions and outlook.---} Quadrature measurements and generation of squeezed states are very important in various fields, including quantum sensing, quantum optics, quantum information, and nanomechanics. The protocol for generation of squeezed states that we presented in this paper makes use of a very basic resource, a two level system with control and measurement. This aspect makes it attractive from a fundamental point of view and at the same time amenable to experimental implementations. Future work will address theoretical aspects of optimization of this protocol for optimal squeezing and tests of the experimental implementation.

\textit{Acknowledgements.---} We thank Martin Otto and Ali Yurtalan for preliminary studies of coupling of a mechanical resonator to a flux qubit, and Aashish Clerk and Eyal Buks for useful discussions. We acknowledge support from Gerald Schwartz and Heather Reisman Foundation, NSERC, Ontario Ministry of Research and Innovation, Industry Canada, and the Canadian Microelectronics Corporation. During part of this work, M.C. was supported for one year by the Scientific and Technological Research Council of Turkey and A.L. was supported by an Ontario Early Research Award.

%

\pagebreak
\widetext

\newcommand{\beginsupplement}{%
        \setcounter{page}{1}
        \setcounter{table}{0}
        \renewcommand{\thetable}{S\arabic{table}}%
        \setcounter{figure}{0}
        \renewcommand{\thefigure}{S\arabic{figure}}%
        \setcounter{equation}{0}
        \renewcommand{\theequation}{S\arabic{equation}}
}
\beginsupplement
\renewcommand{\bibnumfmt}[1]{[S#1]}
\renewcommand{\citenumfont}[1]{S#1}

%
%
%
%
%
%

\begin{center}
\textbf{\large Supplementary material: Quadrature readout and generation of squeezed states of a harmonic oscillator using a qubit-based indirect measurement}
\end{center}

\section{Derivation of the expressions for probability, quadrature average and variance \label{sec:si:analyt:pro:exp:var'}}
In this section we present a derivation of the expressions for $P_{(s-n,n)}$, $\langle I \rangle$, and $\langle I^2 \rangle$ in Eqs. (1)--(3) of the main text. The probability to obtain $n$ results $r=1$ in a series of $s$ measurement is $P_{(s-n,n)} = \bra{\alpha_0}D^{\dagger (s-n)}_{g}D^{\dagger n}_{e}D^{n}_{e}
D^{(s-n)}_{g}\ket{\alpha_0}$, with $\ket{\alpha_0}$ the initial harmonic oscillator (HO) state, taken to be a coherent state of complex amplitude $\alpha_0$. We have $P_{(s-n,n)} = i^{s-2n} \bra{\alpha_0}D^{2n}_{e}
D^{2(s-n)}_{g}\ket{\alpha_0}$, where we used $\left[ D_e, D_g \right]=0$, $D_g^\dagger = i D_g$, and $D_e^\dagger = -i D_e$. Using the binomial theorem, this expression can be expanded as
\begin{equation}
P_{(s-n,n)} =  \frac{i^{(s-2n)}}{2^{2s}} \sum_{k=0}^{2(s-n)} \sum_{\ell=0}^{2n}  {2(s-n)\choose k}{2n\choose \ell} i^{\ell-k}
\bra{\alpha_0}D^{2(s-k-\ell)}\ket{\alpha_0}
\label{equ:P(s-n,n):SI}.
\end{equation}
When the last factor in (\ref{equ:P(s-n,n):SI}) in the bra-ket notation  is expanded further by employing the formulas $D^n \left|\alpha\right\rangle = e^{in\phi\Re\{\alpha\}}\left|(\alpha + n i\phi)\right\rangle$ and $\left\langle\alpha_{i} |\alpha_{j} \right\rangle = \exp\left( \alpha^*_{i}\alpha_{j} -\frac{\left|\alpha_{i}\right|^2}{2} -\frac{\left|\alpha_{j}\right|^2}{2}\right)$~\cite{walls_1995_1SI}, the expression for $P_{(s-n,n)}$ given in Eq. (1) of the main text is obtained.

The quadrature average $\left\langle I\right\rangle_{(s-n,n)} = \bra{\alpha_0}D^{\dagger (s-n)}_{g}D^{\dagger n}_{e}(a+a^\dagger)D^{n}_{e}
D^{(s-n)}_{g}\ket{\alpha_0}/P_{(s-n,n)} $. Using the relations above and $\left[a+a^\dagger , D_g \right] = \left[a+a^\dagger , D_e \right] = 0$, we obtain $\left\langle I\right\rangle_{(s-n,n)} = i^{s-2n}\bra{\alpha_0} (a+a^\dagger)D^{2n}_{e}
D^{2(s-n)}_{g}\ket{\alpha_0}/P_{(s-n,n)}$. This is expanded as
\begin{equation}
\left\langle I\right\rangle_{(s-n,n)} P_{(s-n,n)}
=\frac{\big(-1\big)^{\frac{s}{2}-n}}{2^{2s}}
\sum_{k=0}^{2(s-n)} \sum_{\ell=0}^{2n}  {2(s-n)\choose k}{2n\choose \ell} i^{\ell-k}\bra{\alpha_0} \big(a+a^\dagger\big)D^{2(s-k-\ell)}\ket{\alpha_0}.
\label{equ:<I>(s-n,n):SI}
\end{equation}
After  using  $\left\langle \alpha_{i}\right|I\left|\alpha_{j}\right\rangle =  \left(\alpha_{j} + \alpha^*_{i} \right) \left\langle \alpha_{i}\right.\left|\alpha_{j}\right\rangle$, equation (\ref{equ:<I>(s-n,n):SI}) yields the final form of $\left\langle I\right\rangle_{(s-n,n)}$ in the main text.

Similarly, we can expand $\left\langle I^2\right\rangle_{(s-n,n)} = \bra{\alpha_0}D^{\dagger (s-n)}_{g}D^{\dagger n}_{e}(a+a^\dagger)^2D^{n}_{e}
D^{(s-n)}_{g}\ket{\alpha_0}/P_{(s-n,n)}$ as
\begin{equation}
\left\langle I^2\right\rangle_{(s-n,n)} P_{(s-n,n)} = \frac{\big(-1\big)^{\frac{s}{2}-n}}{2^{2s}} \sum_{k=0}^{2(s-n)} \sum_{\ell=0}^{2n}  {2(s-n)\choose k}{2n\choose \ell} i^{\ell-k}
\bra{\alpha_0}\big(a+a^\dagger\big)^2D^{2(s-k-\ell)}\ket{\alpha_0}
\label{equ:<I2>(s-n,n):SI}.
\end{equation}
Using $\left\langle \alpha_{i} \right| I^2 \left|\alpha_{j}\right\rangle =
\left(1 + \left(\alpha_{j} + \alpha_{i}^*\right)^2
\right) \left\langle \alpha_{i} \right| \left.\alpha_{j}\right\rangle$, we can obtain the final form of $\left\langle I^2\right\rangle_{(s-n,n)}$  in Eq.(3).

We can establish the following relation between the probability, quadrature average, and quadrature square average:
\begin{equation}
\frac{1}{P_{(s-n,n)}}\frac{d}{d\phi}\bigg(\phi P_{(s-n,n)}\bigg) = \langle I^2\rangle_{(s-n,n)} - 2\Re\{\alpha\} \langle I\rangle_{(s-n,n)}
\label{equ:recurrence:SI}.
\end{equation}
This relation will be used in the following section.

\section{derivation of an approximate formula for the variance \label{sec:si:approx:varI} }
In this section we derive an expression for the quadrature variance. We assume a starting vacuum state, $\ket{\alpha_0}=\ket{0}$ and we focus on $n=s/2$ ($s$ is taken even), which is the most likely result. We have
\begin{equation}
P_{(\frac{s}{2},\frac{s}{2})} = \sum_{q=0}^{2s}
e^{- 2\phi^{2}\big(s-q\big)^{2}}
R\big(q\big),
\label{equ:(1):SI}
\end{equation}
where
\begin{equation}
R\big(q\big)=\frac{i^{q} }{2^{2s}}\sum_{k=\max\{0,q-s\}}^{\min\{q,s\}}(-1)^{k}{s\choose k}{s\choose q-k}\label{equ:R(q):SI}.
\end{equation}
The function $R(q)$ is maximum at $q=s$ and symmetric around $q=s$ (i.e. $R (s-a) = R(s+a)$ with $a\leq s$). We have:
\begin{equation}
R (s-a) =\frac{(-1)^{-\frac{s-a}{2}}}{2^{2s}}
{}_{2}\mathrm{F}_{1}\bigg(-s,-(s-a);(a+1);-1\bigg) = \frac{1}{2^{2s}} {s\choose \frac{s+a}{2}}
\label{equ:R(s-a):SI:pre}.
\end{equation}
The combinatorial factor ${s\choose \frac{s+a}{2}}$ in (\ref{equ:R(s-a):SI:pre}) is well described by normal approximation \cite{boo:Kenneth2001},
\begin{equation}
R(s-a) \approx
\frac{1}{2^{s}} \sqrt{\frac{2}{s\pi}} e^{-\frac{a^2}{2s}}
\label{equ:R(s-a):SI}
\end{equation}
if $\abs{a}$ is not much larger than $\rt{s}$.
With (\ref{equ:R(s-a):SI}), equation (\ref{equ:(1):SI}) becomes
\begin{equation}
P_{\frac{s}{2},\frac{s}{2}} = \sqrt{\frac{2}{s\pi}}\frac{1}{2^{s}}
 \sum_{b=-\frac{s}{2}}^{\frac{s}{2}} e^{-\beta b^2}
\label{equ:(1):SI:2}
\end{equation}
where $\beta = \big(\frac{2}{s} + 8\phi^2\big)$.
The sum  in (\ref{equ:(1):SI:2}) is well approximated in the limit $s\rightarrow\infty$  by the Jacobi theta function: $\vartheta_{3}\bigg(0,e^{-\beta}\bigg) = \sum_{b=-\infty}^{\infty} e^{-\beta b^2}$ which is approximated as
\[
\vartheta_{3}\bigg(0,e^{-\beta}\bigg) \backsimeq \sqrt{\frac{s\pi}{2}}
\frac{1}{\sqrt{1 + 4\phi^2 s}}
\]for $e^{-\beta}\rightarrow 1$.
The final form of the asymptotic probability becomes
\begin{equation}
P_{\frac{s}{2},\frac{s}{2}} =
\frac{1}{2^{s}\sqrt{1 + 4\phi^2 s}}
\label{equ:(1):SI:3}.
\end{equation}
Using Eq.~(\ref{equ:recurrence:SI}) for $n=\frac{s}{2}$ and noting that $\langle I\rangle_{(s/2,s/2)} = 0$, the variance becomes
\begin{equation}
\big(\Delta I\big)^2_{\frac{s}{2},\frac{s}{2}}  =  \left\langle I^2\right\rangle_{\frac{s}{2},\frac{s}{2}} = \frac{1}{P_{\frac{s}{2},\frac{s}{2}}}\frac{d}{d\phi}\bigg(\phi P_{\frac{s}{2},\frac{s}{2}}\bigg)=\frac{1}{1+4\phi^2 s}
\label{equ:DeltaI(s/2,s/2):pre1},
\end{equation}
which is result (4) in the main text.


\section{Models for decoherence}

\subsection{Pure dephasing due to flux noise\label{sec:si:decohere:dephase}}
For a superconducting flux qubit, flux noise is the dominant source of dephasing. In the energy eigenbasis, the Hamiltonian of the qubit is
\begin{equation}
H^\prime_\t{qb} = -\frac{\omega_{ge}+\delta\omega_{ge}(t)}{2}\sigma_z,
\label{equ:Hq:noise}
\end{equation}
where $\omega_{ge}=\sqrt{\Delta^2+\varepsilon^2}$, with $\Delta$ the so-called qubit gap and  $\varepsilon=(2I_p\Phi_0/\hbar)(f-1/2)$, with $f=\Phi/\Phi_0$, where $\Phi$ is the externally applied magnetic flux and $\Phi_0=h/2e$ is the magnetic flux quantum, and $I_p$ the qubit persistent current~\cite{orlando_1999_1}. The term $\delta\omega_{ge}(t)$ is a random component induced by intrinsic fluctuations of magnetic flux.

\begin{figure}[!ht]
	\centering
\includegraphics[width=0.95\linewidth]{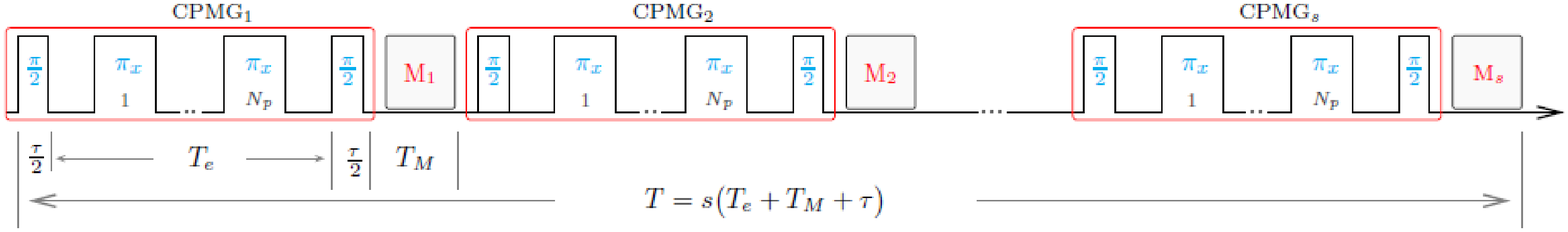}
	\caption{Quadrature detection using the qubit. Each $\mathrm{CPMG}_m$ in this schema is represented  in Fig.~1.a in the main text. \label{fig:QuadDetect:SI}}
\end{figure}

The random component $\delta\omega_{ge}(t)$ in (\ref{equ:Hq:noise}) is a stochastic process, which can be written as
\begin{equation}
\delta\omega_{ge}(t) = \sum_{n=-\infty}^{\infty} a_n e^{i\omega_{n}t}
\label{equ:deltaf:SI}
\end{equation}
where $\omega_{n}=n\times\omega_\mathrm{min}$, with $\omega_\mathrm{min} = \frac{2\pi}{\overline{T}}$, where $\overline{T}$ is a time taken much longer than the duration of the simulated experiment $T$ (see Fig.~\ref{fig:QuadDetect:SI}). The coefficients $a_n$ are taken as random Gaussian variables.
The low frequency noise process $\delta \omega_{ge}(t)$ is characterized by power spectral density (PSD) $S\left(\omega\right) =A_{\omega}/\left|\omega\right|$ where $\sqrt{A_{\omega}} = 2\pi (\varepsilon/\omega_{ge})(I_p/|e|)\sqrt{A_f}$, where the PSD of flux noise is assume to be $A_f/\abs{\omega}$.

We assume $\sqrt{A_f}=10^{-6}$ and $I_{p}=300$ nA, typical for a flux qubit. For simulations of noise trajectories, we restrict the noise frequency to the range $\omega_\mathrm{min}\leq \omega \leq\omega_\mathrm{max}$
where $\omega_\mathrm{min}= 2\pi/ (v\overline{T})$   and $\omega_\mathrm{max}=u N_{p}
2\pi/T_{e}$ with $u$ and $v$ are taken sufficiently large to reflect the relevant time scales and $N_{p}$ is number of pulses.
We construct noise trajectories by using randomly generated complex Fourier coefficients $a_n$, related to the PSD by
$S(\omega) = \frac{\overline{T}}{2\pi} \left\langle\left|a_n\right|^2\right\rangle$.

\subsection{Dissipative effects in the coupled system \label{sec:si:decohere:lindblad}}
We model decoherence of the coupled qubit-harmonic oscillator system using the master equation in Lindblad form:
\begin{eqnarray}
\dot{\rho} = - i\left[\mathcal{H}, \rho\right]
+ \kappa_{\downarrow} \mathcal{D}[a]\rho
+ \kappa_{\uparrow} \mathcal{D}[a^\dagger]\rho
+ \Gamma_{e\rightarrow g} \mathcal{D}[\sigma^{-}]\rho
+ \Gamma_{g\rightarrow e} \mathcal{D}[\sigma^{+}]\rho
+ \Gamma_{\varphi} \mathcal{D}[\sigma_{z}]\rho
\label{equ:lindblad:SI}
\end{eqnarray}
where $\rho(t)$ is the density matrix, $\mathcal{H}$ is the Hamiltonian, $\mathcal{D}[\mathcal{O}]$ is the
damping factor  defined by a mapping
\[\mathcal{D}[\mathcal{O}]\rho =
\mathcal{O} \rho(t) \mathcal{O} ^\dagger -\frac{1}{2}\left( \mathcal{O}^\dagger \mathcal{O} \rho + \rho \mathcal{O}^\dagger \mathcal{O} \right).
\]
In equation (\ref{equ:lindblad:SI}), $\kappa_{\downarrow} = \kappa (1+n_\mathrm{HO})$ and $\kappa_{\uparrow} = \kappa n_\mathrm{HO}$ are the decay rates with average photon number  $n_\mathrm{HO}(\omega_{r}) = 1/\big(\exp(\hbar \omega_{r}/k_{B}T_{HO})-1\big)$  at frequency $\omega_{r}$ and a finite temperature $T_\mathrm{HO}$ ($k_B$ is the Boltzmann constant)
and  $\kappa = \omega_r/ Q$  is the decay rate of the resonator with a quality factor $Q$~\cite{gardiner_2000_1}. Besides, $\Gamma_{e \rightarrow g}$ and
$\Gamma_{g\rightarrow e} = \Gamma_{e\rightarrow g} \exp\left(-\hbar \omega_{ge}/k_B T_\mathrm{qb}\right)$
 are decay rates for the qubit at a finite temperature $T_{qb}$ with
$\Gamma_{e \rightarrow g}+\Gamma_{g \rightarrow e}=1/T_{1,\mathrm{qb}}$ and $T_{1,\mathrm{qb}}$ is the  relaxation time;
$\Gamma_{\varphi}$ is the pure dephasing rate.

The Hamiltonian $\mathcal{H}$ that governs the master equation (\ref{equ:lindblad:SI}) is given by
\begin{equation}
\mathcal{H} = \sum_{i=\{x,y,z\}} \bigg\{ a f_i^*(t) + a^\dagger f_i(t) + f_{c_i}(t) -\frac{1}{2}f_{\varepsilon_i}(t)\bigg\} \sigma_i  +
\frac{A(t)}{2}\frac{\Delta}{\omega_{ge}}\bigg\{
\cos\varphi(t) \sigma_x
-\sin\varphi(t) \sigma_y \bigg\},
\label{equ:H:fluxnoise}
\end{equation}
where
\[\renewcommand{\arraystretch}{1.6}
\begin{array}{lll}
f_z = g\frac{\varepsilon}{\omega_{ge}} e^{i\omega_r t}, & f_y = - g\frac{\Delta}{\omega_{ge}} e^{i\omega_r t} \sin\left(\omega_{ge}t\right), & f_x = -g\frac{\Delta}{\omega_{ge}} e^{i\omega_r t} \cos\left(\omega_{ge}t\right),\\
f_{c_z} =-A \frac{\varepsilon}{\omega_{ge}} \cos\left(\omega_{ge}t+\varphi\right), &
f_{c_y} = \frac{A}{2}\frac{\Delta}{\omega_{ge}}\sin\left(2\omega_{ge}t+\varphi\right), &
f_{c_x} = \frac{A}{2}\frac{\Delta}{\omega_{ge}}\cos\left(2\omega_{ge}t+\varphi\right),\\
f_{\varepsilon_z}=\delta\omega_{ge}(t),
&f_{\varepsilon_y}=\delta\omega_{ge}(t)\frac{\Delta}{\varepsilon}\sin(\omega_{ge}t),&
f_{\varepsilon_x}=\delta\omega_{ge}(t)\frac{\Delta}{\varepsilon}\cos(\omega_{ge}t)
\end{array}
\]
with  $\delta\omega_{ge}(t)$ is given in (\ref{equ:deltaf:SI}).
In this study, we omitted the influence of the last term in (\ref{equ:lindblad:SI}) by setting $\Gamma_{\varphi}=0$,
 since the flux noise is the dominant source of dephasing and its contribution is embedded into the Hamiltonian (\ref{equ:H:fluxnoise}).

\section{Derivation of conditional evolution with pure dephasing \label{sec:si:dephase}}
With pure dephasing, a random phase $\tilde{\phi}_m$ is added to a qubit superposition, which has a different value for each repetition from $m=1$ to $s$. The noise is drawn from a proper distribution corresponding to the noise spectrum and taking into account the noise modulation due to the CPMG sequence (Fig.~\ref{fig:QuadDetect:SI}).
Using the projection operator $D_{g(e)}=-\frac{1}{2}(D-(+)e^{i\tilde{\phi}}iD^\dag)$ and following the procedure in the main text, one can obtain pure state conditioned on measurement result $\mathbf{r}$:
\begin{equation}
\ket{\alpha_\mathbf{r}} =   \frac{(-1)^s}{2^s}\prod_{m=s}^{1}\sum_{q_m=0}^{1} i^{r_m q_m} e^{i\tilde{\phi}_m q_m} D^{1-2q_m}\ket{\alpha_0}=
\frac{(-1)^s}{2^s} \sum_{\mathbf{q}=0}^{1} i^{\mathbf{r}\cdot \mathbf{q}} e^{i\mathbf{\tilde{\phi}}\cdot\mathbf{q}} D^{s-2t}\ket{\alpha_0}
\label{equ:purestate:SI}
\end{equation}where $\ket{\alpha_0}$ is the initial state,
$\mathbf{q}$ is a vector of length $s$ with components $q_i=0,1$ ($i=\overline{1,s}$), $\mathbf{\tilde{\phi}}$ is the vector of length $s$ formed of the values of the random phase $\tilde{\phi}_i$ ($i=\overline{1,s}$),and
$t=\sum_{j=1}^{s} q_{j}$.
Consequently, the density matrix in the main text is obtained
from (\ref{equ:purestate:SI}) by averaging over the noise process:
\begin{equation}
\rho_\mathbf{r}=\frac{1}{2^{2s}}\sum_{\mathbf{q_1},\mathbf{q_2}}i^{\mathbf{r}\cdot(\mathbf{q_1}-\mathbf{q_2})}
C_{\mathbf{q_1},\mathbf{q_2}}
D^{s-t_1}\ket{\alpha}\bra{\alpha}{D^\dag}^{s-t_2},
\end{equation}
where $C_{\mathbf{q_1},\mathbf{q_2}} = \left\langle e^{i\mathbf{\tilde{\phi}}\cdot(\mathbf{q_1}-\mathbf{q_2})} \right\rangle=\exp{\left( -\frac{1}{2} (\mathbf{q_1}-\mathbf{q_2})  \mathbf{\widetilde{W}} (\mathbf{q_1}-\mathbf{q_2})\right)}$ with
$\mathbf{\widetilde{W}}$ the correlation matrix for the noise $\tilde{\phi}_1,\tilde{\phi}_2,\ldots,\tilde{\phi}_s$.

The correlation matrix can be written as
\begin{equation}
\widetilde{W}_{ij} =
\int_{t_i}^{t_{i} + T_e}dt \chi(t)\int_{t_j}^{t_{j} + T_e}dt'\chi(t')\overline{\big(\delta\omega_{ge}(t)\delta\omega_{ge}(t')\big)}  = \int_{-\infty}^{\infty} d\omega S(\omega)
e^{i\omega(t_j-t_i)} \big|\widetilde{\chi}(\omega)\big|^2
\label{equ:correlation:SI}
\end{equation}
where $\widetilde{\chi}$ is Fourier transform of $\chi(t)$ depicted in Fig.~1.(a) in the main text.  We find that to a good approximation, the correlation matrix (\ref{equ:correlation:SI}) is diagonal, with elements
\begin{equation}
\widetilde{W}_{ii} \approx 0.424 \times A_{\omega} \bigg(\frac{\pi}{\omega_r}\bigg)^2  N_{p},\;\;\;\;\forall\;i=1,\ldots,s
\label{equ:correlation:Wii}.
\end{equation}

%
%

\end{document}